\documentclass[12pt]{article}
\usepackage{epsfig}
\begin{document}

\begin {center}
{\Large \bf An explanation of Belle states $Z_b(10610)$ and
$Z_b(10650)$}
\vskip 5mm
{D V Bugg \footnote{email: d.bugg@rl.ac.uk}\\
{\normalsize  \it Queen Mary, University of London, London E1\,4NS,
UK} \\
[3mm]}
\end {center}
\date{\today}

\begin{abstract}
Belle report data on $\Upsilon (5S) \to \Upsilon(1S,2S,3S)\pi ^+\pi^-$
and $\chi_b(1P,2P)\pi ^+\pi ^-$ ; they observe peaks in $\Upsilon
\pi^\pm$ and $\chi_b\pi ^\pm$ consistent with $J^P=1^+$.
They interpret the peaks as molecular states $Z_b(10608)$ and
$Z_b(10653)$.
Their masses are just above $\bar B B^*$ and $\bar B^*
B^*$ thresholds at 10604.6 and 10650.2 MeV.
An explanation in terms of cusps at these thresholds is presented here.
The product of the rising phase space for $\bar B B^*$ and
$\bar B^* B^*$ with the cusps creates peaks a few MeV higher in
$\bar B B^*$ and $\bar B^* B^*$, and these peaks can de-excite to
$\pi ^+\pi ^-\Upsilon (1S,2S,3S)$ and $\pi ^+\pi ^-\chi_b(1P,2P)$.

\end{abstract}

Belle \cite {Idachi} report peaks in $\pi ^\pm\Upsilon (1S,2S,3S)$ and
$\pi ^\pm \chi_b(1P,2P)$ with masses and widths:
\begin {eqnarray}
M[Z_b(10610)] = 10608.4 \pm 2.0 \,{\rm MeV} &\Gamma = 15.6 \pm 2.5
\, {\rm MeV} \\
M[Z_b(10650)] = 10653.2 \pm 1.5 \,{\rm MeV} &\Gamma = 14.4 \pm
\, 3.2 {\rm MeV}.
\end {eqnarray}
Angular distributions are consistent with $J^P = 1^+$ in both cases.

Thresholds for $B^*\bar B$ and $B^*\bar B^*$ are at 10604.6 MeV and
10650.2  MeV and have the same mass separation
as the two $Z_b$ peaks within experimental errors.
This suggests strongly that the $Z_b$ peaks are associated with those
thresholds.
There is an explanation how this can arise via cusps at
the $B^*\bar B$ and $B^* \bar B^*$ thresholds.

Consider $B^*\bar B^*$ elastic scattering first.
The imaginary part of the T-matrix for $B^*\bar B^* \to B^*\bar B^*$
S-wave scattering is
\begin {equation}
{\rm Im}\, T \propto g^2\rho (s),
\end {equation}
where $\rho$ is the phase space for $B^*\bar B^*$ and $g$ is the
coupling constant to this channel.
Analyticity requires a real part to the amplitude \cite {Sync}
\begin {equation}
{\rm Re}\, T \propto \frac {1}{\pi } P
\int _{s_{thr}} ^{M^2[\Upsilon (5S)]}
\frac {ds' g^2(s') \rho (s')}{s' - s},
\end {equation}
where $s_{thr}$ is mass squared at threshold;
P denotes the principal value integral.
For $B^*\bar B^*$, the S-wave phase space is
\begin {equation}
\rho (s') = \frac {2k}{\sqrt {s'}} FF(s'),
\end {equation}
where $k$ is the momentum of each $B^*$ in the $B^*\bar B^*$ rest frame
and $FF(s')$ is a form factor.
There are further form factors  for $\Upsilon (5S) \to B^*\bar B^*$,
but these change little over the cusps and will be neglected.
It is important to be aware that this cusp effect is not an `optional
extra'. 
It is a basic requirement at the opening of any new channel, but has
been ignored in the majority of calculations of `molecules'.
It is also known as the `coupled channel effect'. 

\begin{figure}[htb]
\begin{center} \vskip -52mm
\epsfig{file=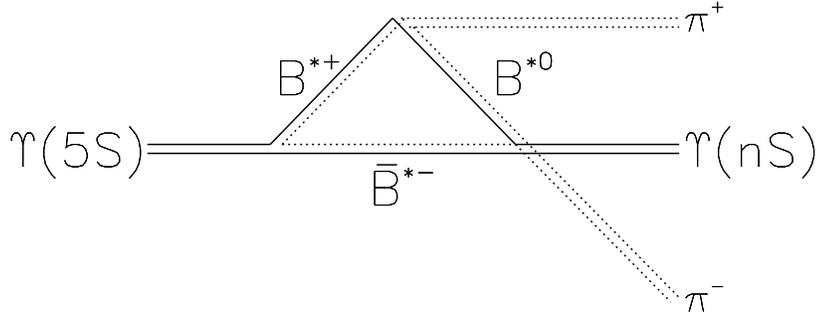,width=15cm}
\vskip -52mm
\caption {One diagram for $\Upsilon (5S) \to \pi^+ B^*\bar B^*$
followed by de-exitation $B^*\bar B^* \to \pi^- \Upsilon (nS)$.
Full lines represent $b$ and $\bar b$, dashed lines light quarks.}
\end{center}
\end{figure}
Let us now turn to the process
$\Upsilon (5S) \to \pi ^+\pi ^-\Upsilon (nS)$ shown in Fig. 1.
It goes via re-arrangement of the four quarks in $B^*\bar B^*$.
This process has the same analytic structure as
$B^*\bar B^* \to B^*\bar B^*$, since that amplitude obviously
factorises.
Let us consider first $B^*\bar B^* \to \Upsilon (nS)\pi$ via $\pi$ exchange.
Heavier exchanges will be discussed later.  
The momentum transfer in the production is quite large and
varies little over the cusp, though the phase space for the intermediate
$B^*$ is included.
For free $B^*$, pion decay is forbidden by the small mass
difference between $B^*$ and $B$.
However, for the triangle graph shown in Fig. 1, intermediate $B^*$
are off-shell, allowing $\pi$ exchange in $B^*\bar B$ scattering at the
right-hand vertex.
Suppose the form factor is a Gaussian
\begin {equation}
FF = \exp \,(-k^2R^2/3)
\end {equation}
with $R = 1.41$ fm, corresponding to the mass of the pion, though the
exponential may be an approximation.

\begin{figure}[htb]
\begin{center}
\vskip -16mm
\epsfig{file=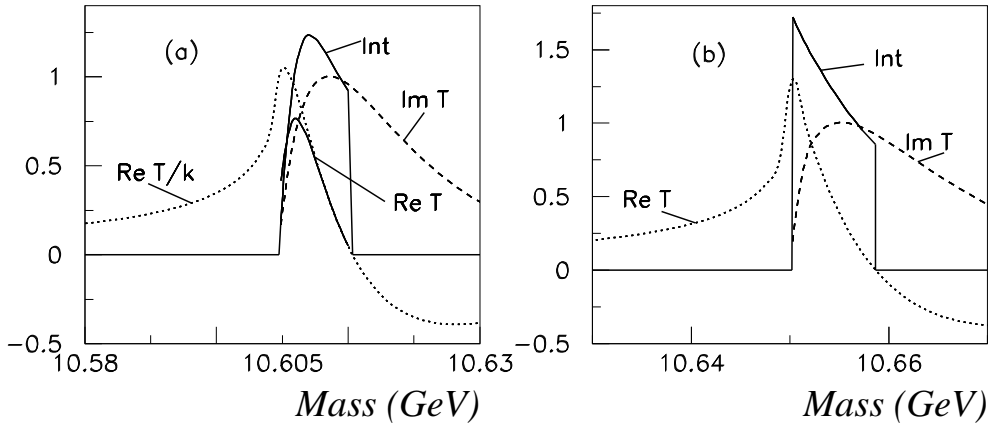,width=15cm}
\vskip -8mm
\caption {${\rm Im} T$, ${\rm Re}\, T$ or $T/k$ (dotted) and
$|T|^2$ (full curve) for (a) $B^*\bar B \to B^*\bar B$ and (b)
$B^*\bar B^* \to B^*\bar B^*$; Int stands for Intensity.
On Fig. 2(a), ${\rm Re} T$ is also shown  by a full curve.}
\end{center}
\end{figure}
Fig. 2(b) shows ${\rm Im}\, T$ as the dashed curve,
normalised to 1 at its peak.
Then ${\rm Re}\, T$ , shown as the dotted curve with the same
normalisation, has a sharp cusp exactly at the $B^*\bar B^*$ threshold.
${\rm Im}\, T$ peaks at a mass of 10656 MeV.
${\rm Re}\, T$ goes negative just above this mass.
For higher masses, there is repulsion between $B^*$ and $\bar B^*$;
this will keep them apart and suppress de-excitation to
$\pi \Upsilon (nS)$, where $n = 1-3$.
As a result, the intensity $|T|^2$ for $\Upsilon (5S) \to \pi ^+\pi ^-
\Upsilon (nS)$ is proportional to the full curve.
It rises at the threshold 10650.2 MeV to the scattering length given
by the cusp and switches off at 10658.7 MeV.
Folding in the 5.2 MeV experimental resolution quoted by Belle gives
a somewhat rounded line-shape with a full-width at half-height of
10.0 MeV, rather less than the Belle value of $14.4 \pm 3.2$ MeV.

Consider next $B^*\bar B$ scattering via a diagram like Fig. 1 except
for a $\bar B^-$ running from left- to right-hand vertices.
This requires a P-wave interaction and
\begin {equation}
\rho (B^* \bar B) = k^3 \exp(- k^2 R^2)/(1 + k^2 R^2),
\end {equation}
assuming a centrifugal barrier with the same radius as the exponential;
this is an approximation where the centrifugal barrier is replaced by
an equivalent square barrier.
It is necessary to preserve the P-wave form for the amplitude at
threshold and the dispersion integral becomes
\begin {equation}
{\rm Re}\, T \propto  \frac {k}{\pi} P
\int _{s_{thr}} ^{M^2[\Upsilon (5S)]}
\frac {ds' g^2(s') \rho (s')/k'}{s' - s}.
\end {equation}
Fig. 2(a) shows the corresponding line-shape for ${\rm Re}\, T/k$
as the dotted curve; as on Fig. 2(b), it peaks at threshold.
When it is multiplied by the  factor $k$ in Eq. (8),
the line-shapes for ${\rm Re }\, T$ and $|T|^2$ are shown by the full
curves and ${\rm Im}\, T$ is shown by the dashed curve.
In this case, the threshold is at 10604.6 MeV and the $B^* \bar B$
interaction becomes repulsive at 10613.9 MeV.
The peak is centred at 10609.2 MeV and its full width becomes 9.5 MeV
when folded with experimental resolution, compared with the
experimental value $15.6 \pm 2.5$ MeV.

There are two key points in favour of the cusp mechanism.
The first is that it predicts almost the same widths for $Z_b(10610)$ and
$Z_b(10650)$, in agreement with data. 
In the molecular picture, these widths have no obvious relation.
Secondly, the measured branching ratios of $\Upsilon (1S,2S,3S)\pi \pi$
sum to $\sim 2\%$ of the total width of $\Upsilon (10860)$, hence accounting
for only a 2 MeV width for each peak \cite {PDG}. 
There is a factor 7 discrepancy with experimental values. 

Let us now consider heavier exchanges, for example due to $\sigma$.
A full calculation is plagued by unknowns.
Firstly, it is unclear whether the $\sigma$ will couple via
the $\sigma$ pole (as in BES II data for $J/\Psi \to \omega \pi \pi$
\cite {BESII}) or as in elastic scattering, where the Adler zero
appears in the numerator of the amplitude.
Next, it is necessary to fold the radius of interaction (as a function
of $\sigma$ mass) with the folded radius of two $B^*$ (or $\bar B B^*$).
A calculation gives an RMS radius of $\sim 0.65$ fm after
this folding procedure. 
The cut-off at high $Z_b$ mass varies with the folded mass and is rounded 
off both below and above the sharp cut-off as shown below in Fig. 3. 
However, the key point is that, classically, the cross section for
each de-excitation process, e.g. $\bar B^* B^* \to \Upsilon (nS)\pi$,
is proportional to the mean $<R^2>$.
As a result $\pi$ exchange plays a strong role.  

This procedure will be reversed here, deducing the RMS radius from
the widths of $Z_b(10610)$ and $Z_b(10650)$ and asking whether this is 
reasonable. 
Even then, there is a minor problem that Belle have assumed these two peaks
are resonant and included interferences between them and have also included
interference with a non-resonant background. 
The two most prominent signals in their data are for $Z_b(10610)$ and
$Z_b(10650) \to \Upsilon (2S,3S)\pi \pi$, see their Fig. 5. 
Reading data directly from these graphs, widths come out $17\%$ higher than 
their values.
This is a small discrepancy, but systematic and probably arises from
their inclusion of interferences between the two peaks. 
Using my widths, $\sqrt {<R^2>}$ comes out to be 0.95 fm.
In view of the factor $R^2$ favouring $\pi$ exchange, this appears 
reasonable within $\pm$ 0.1 fm, i.e. $\sim 10\%$.

\begin{figure}[htb]
\begin{center}
\vskip -16mm
\epsfig{file=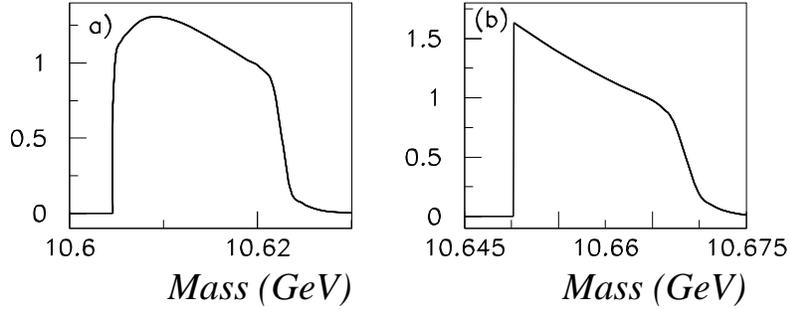,width=12cm}
\vskip -8mm
\caption {Line-shapes of the cusps using a radius of 0.95 fm, adjusted to 
give an optimum fit to Belle data.}
\end{center}
\end{figure}
Fig. 3 shows resulting line-shapes for this radius of interaction. 
The sharp cut-offs of Fig. 2 are slightly rounded off, but are still
distinctive.
With further data of higher statistics (and perhaps improved
mass resolution), this line-shape is open to experimental test.
In work concurrent with the present paper, Danilkin, Orlovsky and
Simonov discuss the general formalism of heavy quarkonia
\cite {Simonov} and show results for $Z_b(10610)$ and $Z_b(10650)$;
they use a smaller radius of interaction and therefore find wider
cusps which are inconsistent with the data. 

It is possible that the $Z_b$ states are actually resonant.
Cleven, Guo, Hanhart and Mei$\beta$ner assume the molecular state is
resonant and close to threshold; they adjust coupling parameters to 
reproduce the data on $h_b(2P)$ final states \cite {Hanhart}.
They include the cusp effect and arrive at quite different line-shapes
to those presented here. 
Their essential feature is a sharp cusp right at threshold.   

\begin{figure}[htb]
\begin{center} \vskip -16mm
\epsfig{file=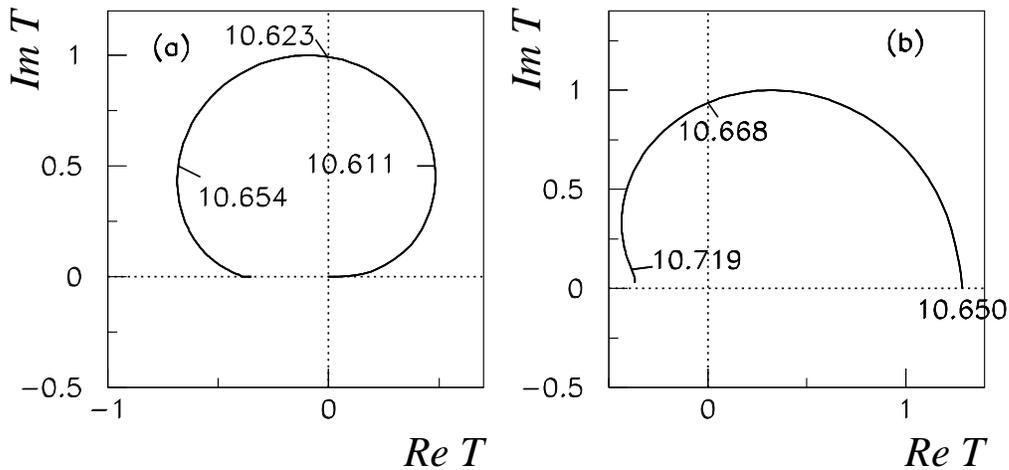,width=15cm}
\vskip -8mm
\caption {Argand diagrams for (a) $B^*\bar B \to B^*\bar B $,
(b) $B^*\bar B^* \to B^*\bar B^*$; numbers are masses in GeV.}
\end{center}
\end{figure}
Figs. 4(a) and (b) show Argand diagrams for both $Z_b$ derived from
Fig. 3; masses are marked in GeV.
In Fig. 4(b) there is no pole, because of the $s$-dependence around the 
curve.
Above a mass where the amplitude goes negative, the variation
with $s$ slows down rapidly.
However, there is a possibility that meson exchanges will
dynamically generate a resonant pole.
The evaluation of the cusp made here is exactly equivalent to evaluating the
first order loop diagram for meson exchanges.
Solving the Bethe-Salpeter equation would include loop diagrams of all
orders and could dynamically generate a resonance.
The more favourable situation is for the loop driven by $\bar B B^*$, 
because of the P-wave factor at threshold.
For Fig. 4(a), if the data are fitted with a Breit-Wigner resonance with
$s$-dependent width, there is a pole at $10.623 \pm 0.005$ MeV.
It has a full width at half-height of 44 MeV and is centred at 10632 MeV.
Note that the loop is really a property of the $\bar B B^*$ channel.
If there is an isospin $I=0$  $\chi_{b1}(P)$ state near this threshold, 
it is likely to be attracted towards the $\bar B^*$ threshold.
There is also the possibility that the $Z_B(10610)$ becomes an exotic 
resonance, but that needs to be proved.

There is one detail concerning decays to $\pi ^+\pi ^-\Upsilon (nS)$.
For this case, both $\pi$ are produced in a relative S-wave and are
fitted with the $\sigma$ amplitude.
To form the final states $\pi ^+ \pi ^- \chi_b(1P,2P)$ observed by
Belle, one pion must be in a P-wave.
This is consistent with the observation of decays to
$\sigma_{L=1}$, where $L$ is the orbital angular momentum of the
$\sigma$ with respect to $\chi_b$ states.
A further detail is that pion production in the step $B^*\bar B^*$
or $B^*\bar B \to \pi \chi_b$ is likely to flip the spin of the
$\chi_b$ final state.
This can account for a phase difference with respect to
$\pi ^+\pi ^-\Upsilon (nS)$ of $\sim 180^\circ$, as recorded
by Belle.
Because of nodes in radial wave functions,
significant differences are to be expected between Dalitz plots for
$\pi ^+\pi \Upsilon (nS)$ with different $n$ values.
There has been earlier discussion of a possible hybrid  in the mass
range 10.4--10.8 GeV on the basis of variations of Dalitz plot for 
$\Upsilon \pi \pi$ with different $n$ values \cite {hybrid}.

In conclusion, threshold cusps at $B^*\bar B$ and $B^*\bar B^*$
thresholds are capable of explaining the peaks observed by Belle.
The predicted line-shapes are distinctive, and with higher statistics
can be used to test the present hypothesis. 
The width predicted for these cusps requires some exchange of heavier 
mesons than pions between $B^*$ and both $\bar B$ and $\bar B^*$.
The cusp mechanism has the virtue of predicting closely similar widths 
for both $Z_b$ peaks.
The experimental width of the peaks is a factor 7 larger than that
for $Z_b \to \Upsilon (1S,2S,3S)\pi \pi$, but is reproduced by the
cusp mechanism with a reasonable radius of interaction.

\end {document}